\documentclass[10pt,letterpaper,twocolumn]{article}
\usepackage[english]{babel}
\usepackage{graphicx}

\newcommand{\alt}{\mathbin{\lower 3pt\hbox
   {$\rlap{\raise 5pt\hbox{$\char'074$}}\mathchar"7218$}}}
\newcommand{\agt}{\mathbin{\lower 3pt\hbox
   {$\rlap{\raise 5pt\hbox{$\char'076$}}\mathchar"7218$}}}

\begin{document}

\setcounter{footnote}{0}
\setcounter{equation}{0}
\setcounter{figure}{0}
\setcounter{table}{0}

\title{\large\bf Is Fermi liquid topologically protected?
 }

\author{ \small I. M. Suslov \\
\small  P.L.Kapitza Institute for Physical Problems,
119334 Moscow, Russia \\
{}\\
\parbox{150mm}{ \footnotesize \, The book by Volovik
\cite{1} contains the argument, which can be considered
as the topological proof of the Luttinger theorem. The Green
function of the ideal Fermi gas has a pole in the $(E,|p|)$
plane (where $E$ and $p$ are  energy and momentum). This
pole is considered
to be analogous to a vortex in liquid
helium. Since a vortex is topologically stable against
restricted
perturbations, one can include interaction adiabatically (as a
succession of small perturbations), and observe
transformation of the  Fermi gas pole to the Fermi
liquid pole. In this
%
argument, the topological stability arises already on the level
of the ideal Fermi gas, which is in conflict with its Cooper
instability.  We discuss the origin of this controversy.  } }

\date{}
\maketitle

\textwidth 6.4 in
\textheight 8.5 in

\setcounter{footnote}{0}
\setcounter{equation}{0}
\setcounter{figure}{0}
\setcounter{table}{0}

The book by Volovik \cite{1} contains the argument, which can be
qualified \cite{2} as the topological proof of the Luttinger
theorem \cite{3}. Below we discuss some controversies related with
this argument.

\vspace*{2mm}

The argument goes as follows.  The Green function of the ideal
Fermi gas has a pole in the $(E,|p|)$ plane (where $E$ and $p$
are  energy and momentum).
If we consider the closed contour around this pole, then
the change of the phase along this contour is $2\pi$, which is
considered as the topological invariant. The same topological
invariant exists in the case of a vortex in liquid helium.
%
%
On this ground, the pole is considered to be topologically
equivalent to the vortex. The vortex is a topologically
stable object:  you can do nothing with it, if you have not at
hand the energy comparable with the vortex binding energy. Hence,
one can include interaction adiabatically (as a succession of
small perturbations) and observe transformation of the Fermi gas
pole to the Fermi liquid pole. Since only existence of the pole
is topologically protected, its position and residue can be
changed, leading to the Fermi liquid renormalization of
parameters. As a result, one reproduces all essential statements
of the Luttinger theorem \cite{3}, which was in fact established
by Landau in his Fermi liquid theory \cite{4}.

\vspace*{3mm}

One can easily suspect that not all is good with this argument.
According to it,  the topological  stability arises already on
the level of the ideal Fermi gas; however, the ideal Fermi
gas posseses the well-known Cooper instability. As a result,
one is staying before a choice: either something is wrong
with this argument, or  "topological stability" is
physically irrelevant in this context. In fact, both things
are true.

\vspace*{3mm}

If references to topology are put aside, the logical mistake is
very simple. The pole and the vortex have one common property:
a change of the phase  is $2\pi$ along the embracing
contour.  However, it does not mean that all other properties are
analogous. In fact, the analogy between the pole and the vortex is
absolutely incorrect. The vortex is a quantized object, while the
pole is not subjected to any quantization: the residue in it can
be changed continuously\,\footnote{\,The helium vortex can be
associated with the pole in the behavior of the order parameter,
but the residue in this pole is a fixed number for a given
temperature. }.
The vortex cannot be divided into two parts, and one cannot nip
off a small piece of it. The latter is easily possible in the
case of the pole: by an arbitrary small perturbation one can nip
off a small piece  (i.e. a pole with a small residue) and
shift it to a small distance. By another small perturbation one
can shift this piece a little more. By the next small
perturbation one can nip off another small piece, and so on. By a
succession of small perturbations the pole can be transformed to
very different configurations of singularities.  Nothing of the
kind is possible with the vortex\,\footnote{\,The topological
protection from perturbations is a serious thing, which is
vividly discussed in the context of quantum computing.  The main
idea is to reduce a system to a set of the quatized objects;
continuous objects are not suitable for it. }.  In the case of
the vortex, "the topological stability" is a property, which is
essentially stronger that the usual stability in the Lyapunov
sense (i.e. stability against infinitely small perturbations of
the general form): it is stability against restricted, but finite
perturbations.

\vspace*{3mm}

The given considerations can be illustrated on the example of
the Anderson model, describing the movement of an electron in
the random potential,
$$
J(\Psi_{n+1}+\Psi_{n-1})+V_n \Psi_n = E \Psi_n \,,
\eqno(1)
$$
and corresponding to discretization of the usual Schroedinger
equation ($J$ is the hopping integral, $E$ is the energy,
$\Psi_n$ is a wave function at the site $n$).  If $J=0$ and
all $V_n$ are equal to $V_0$, then the Green function of (1) has
a simple pole at $E=V_0$. If the model of an alloy is considered,
when $V_n=V_0$ with probability $1-p$ and $V_n=V_1$ with
probability $p$, then the pole is splitted into two poles at
$E=V_0$ and $E=V_1$. It can be done for arbitrary small $p$ and
$V_1-V_0$, which corresponds to an arbitrary small concentration
of arbitrary weak inpurities and is admitted as a small
perturbation according to all physical criteria. If $V_n$ have
the Gaussian distribution with the arbitrary small width, the
pole transforms to the infinite cut\,\footnote{\,The cut is
infinite, because $V_n$ are unbounded for the Gaussian
distribution. If $V_n$ are restricted, then the cut will be
finite. }.

The latter example reveals one more qualitative difference
between the pole and the vortex. The vortex obeys the binding
energy, which provides  rigidity against attempts to deform it.
No energy is associated with the pole and it has not rigidity
against extention.
The given illustration does not
correspond to the book \cite{1} exactly: it deals with the pole
in the plane $({\rm Re}\, E, \,{\rm Im} \,E)$, and not
$(E,\,|p|)$.  However, it is a matter of principle, that
a pole behaves completely differently from a vortex.

\vspace{3mm}

What about references to topology and the topological invariant?
The clear answer can be given for the case of the pole in the
complex $E$ plane, using  the spectral representation of the
Green function
$$
G_E(x,x')=\sum_s \frac{\Psi_s(x)\Psi^*_s(x')}{E-\epsilon_s} \,,
$$
where $\Psi_s(x)$ and $\epsilon_s$ are eigenfunctions and
eigenvalues of the Hamiltonian.  Let begin with some simple
situation like $J=0$ and $V_n=0$ in (1). Then all $\epsilon_s$
are equal, and we have a simple pole in the complex $E$ plane,
obeying the same topological invariant as in
\cite{1}: the change of the phase is $2\pi$ along the embracing
contour. For the sake of rigorousness, the contour should be
chosen sufficiently large.  The inclusion of finite $J$, $V_n$
and other factors transforms the pole to a certain configuration
of singularities. This set of singularities remains inside the
contour, if
added perturbation is restricted, and will
be looking as a simple pole from viewpoint of the infinitely
remote contour.  This is indeed the topological property, which
cannot be spoiled by a restricted perturbation.  However, this
topological property does not protect nobody from anything.  The
model (1) with $V_n=0$ describes the ideal Fermi gas in the
finite band; in the result of different perturbations it can be
transformed to Fermi liquid, Luttinger liquid, superconductor,
Anderson dielectric, Mott dielectric, Peierls dielectric, and so
on. All these things can be realized under existence of the same
topological invariant.  In principle, one can easily believe that
the topological classification allows to distinguish all
essentially different things, but such classification should
include not one, but several topological invariants.

The situation is analogous for the case considered in the book
\cite{1}. There is the only difference, that momentum is
considered as a good quantum number, so $\epsilon_s$ are related
with the allowed values $p_s$ of momentum, while the embracing
contour is chosen in the half-plane $(E,\,|p|)$.


\vspace{3mm}

It should be clear from above that references to topology
should be made with great precaution. It set other questions in
respect to  content of the book \cite{1}. The main thing is
that axiomatics of
the topology
science not always corresponds to physical applications.
The topological invariant is unchangeable by definition
in the topology science.  It does not mean that topologists
naively believe that topology cannot be changed;
it is  simply convenient for them to fix a certain topology and
look what can be done under such restriction. As for physics,
there is no the conservation law for topological invariants.
Sometimes the topological invariant  can be really fixed by the
external conditions (as in the case of different winding
numbers), but sometimes it can easily change under continuous
variation of physical parameters (as in the case of the Lifshitz
topological transitions \cite{5}).  In simple words, the roll can
be made from rubber or plasticine.  The rubber roll is
topologically stable: if one grip it in hand to make a ball from
it, a small gap still remains and the deformed roll remains to
be a roll topologically; and it restores its initial form, when a
pressure is removed.  As for the plasticine roll, there is no
problem to make a ball from it, and there is no topological
stability.

\vspace{3mm}

A simple existence of the topological invariant does not mean
anything. The first question should be asked, is this topological
invariant really unchangeable. Is a situation of rubber or
plasticine? Unfortunately, such questions are never asked in the
book \cite {1}.

\end{document}